\documentclass[aps,prl,reprint,superscriptaddress,amsmath,amssymb,floatfix,nofootinbib]{revtex4-2}

\usepackage[T1]{fontenc}
\usepackage[utf8]{inputenc}
\usepackage{graphicx}
\usepackage{bm}
\usepackage[colorlinks=true,linkcolor=blue,citecolor=blue,urlcolor=blue]{hyperref}

\newcommand{\EN}{E_{\mathcal N}}
\newcommand{\NSM}{\mathcal N_{\mathrm{SM}}}
\newcommand{\ii}{\mathrm i}
\newcommand{\Ccal}{\mathcal C}

\begin{document}

\title{All-optical switching of continuous-variable entanglement \\ in an absorption-suppressed plasmonic nanodimer}

\author{Elif Ozturk}\email{elifozturk24@hacettepe.edu.tr}
\affiliation{Institute of Nuclear Sciences, Hacettepe University, Ankara 06800, T\"urkiye}
\affiliation{Department of Nanotechnology and Nanomedicine, Graduate School of Science and Engineering, Hacettepe University, Ankara 06800, T\"urkiye}
\author{Mehmet Gunay}\email{mehmetgunay35@gmail.com}
\affiliation{Department of Nanoscience and Nanotechnology, Faculty of Arts and Science, Burdur Mehmet Akif Ersoy University, Burdur 15030, T\"urkiye}
\author{Ramazan Sahin}\email{rsahin@itu.edu.tr}
\affiliation{Department of Astronomy and Space Sciences, Atat\"urk University, Faculty of Science, Erzurum 25050, T\"urkiye}
\affiliation{T\"urkiye National Observatories, TUG, Antalya 07058, T\"urkiye}
\author{Mehmet Emre Tasgin}\email{metasgin@hacettepe.edu.tr}
\affiliation{Institute of Nuclear Sciences, Hacettepe University, Ankara 06800, T\"urkiye}
\affiliation{T\"urkiye National Observatories, TUG, Antalya 07058, T\"urkiye}
\date{\today}

\begin{abstract}
	A subwavelength quantum-photonic circuit element should simultaneously
	generate nonclassical light, suppress plasmonic loss, and remain
	dynamically tunable. We show that an orthogonal plasmonic nanorod dimer
	can satisfy all three requirements. A phase-locked control polarization
	induces plasmonic refractive-index enhancement, driving the probe
	response toward a near-zero-extinction regime while simultaneously
	tuning the local second-harmonic parametric interaction. The resulting
	nonlinear plasmonic source operates in an absorption-suppressed regime
	and enables all-optical control of quantum correlations. We demonstrate
	switchable logarithmic negativity and single-mode nonclassicality,
	establishing a route toward actively tunable quantum-plasmonic circuit
	elements operating well below the diffraction limit.
%
\end{abstract}

\maketitle

{\it Introduction.}--- Continuous-variable entanglement and quadrature squeezing are essential resources for quantum communication, quantum sensing, and optical quantum information processing \cite{horodecki2009,weedbrook2012,braunstein2005,andersen2015,furusawa2011}. A useful integrated source should not only generate these resources, but also switch and tune them after fabrication. Plasmonic nanostructures offer the required extreme mode confinement and large local fields \cite{gramotnev2010,tame2013,kauranen2012}; squeezed surface plasmons have already been observed in metal waveguides \cite{huckPRL2009}. The same confinement, however, makes absorption and weak post-fabrication tunability central obstacles for quantum-plasmonic circuit elements.

Second-harmonic generation (SHG) provides a compact parametric route to nonclassical fluctuations \cite{lugiato1983,walls2008,boyd2020,tasgin2023}. In a metallic nanoantenna, the effective SHG coefficient can arise from surfaces, structural asymmetry, or molecules with nonlinear response, and should be interpreted as a localized-mode nonlinear parameter rather than as an ideal bulk electric-dipole susceptibility \cite{kauranen2012,celebrano2015,boyd2020,Zayats2015,huseyintepe2025}. The advance proposed here is to place this nonlinear process under coherent all-optical control while simultaneously operating at a point of strongly suppressed effective absorption. The control mechanism is the plasmonic analogue of enhancement of the refractive index: a mutually coherent orthogonal polarization modifies the probe polarizability through interference, allowing a large dispersive response with vanishing or negative effective extinction \cite{scully1991,fleischhauer1992,YavuzPRL2005,YavuzPRL2008,panahpour2019,dhama2022,yuce2021}~\footnote{\label{two_optical_modes}The scheme studied in Ref.~\cite{panahpour2019} is more appropriately compared with the atomic-vapor EIR schemes of Refs.~\cite{YavuzPRL2005,YavuzPRL2008}, where the refractive-index enhancement is achieved through the coherent interaction of {\it two optical fields} with similar frequencies rather than through optical--microwave coupling. Those experiments directly demonstrated the large optical phase shifts enabled by the enhanced refractive index. }.

We use this effect to convert a classical index-enhancement knob into a quantum entanglement switch. Such compact and reconfigurable sources of continuous-variable entanglement are particularly relevant for measurement-based quantum computing, where quantum processing is performed through measurements on highly entangled resource states, eliminating the need for strong nonlinear elements within the computational circuit itself~\cite{Furusawa_MBQC}.


\begin{figure}[t]
    \centering
    \includegraphics[width=0.95\columnwidth]{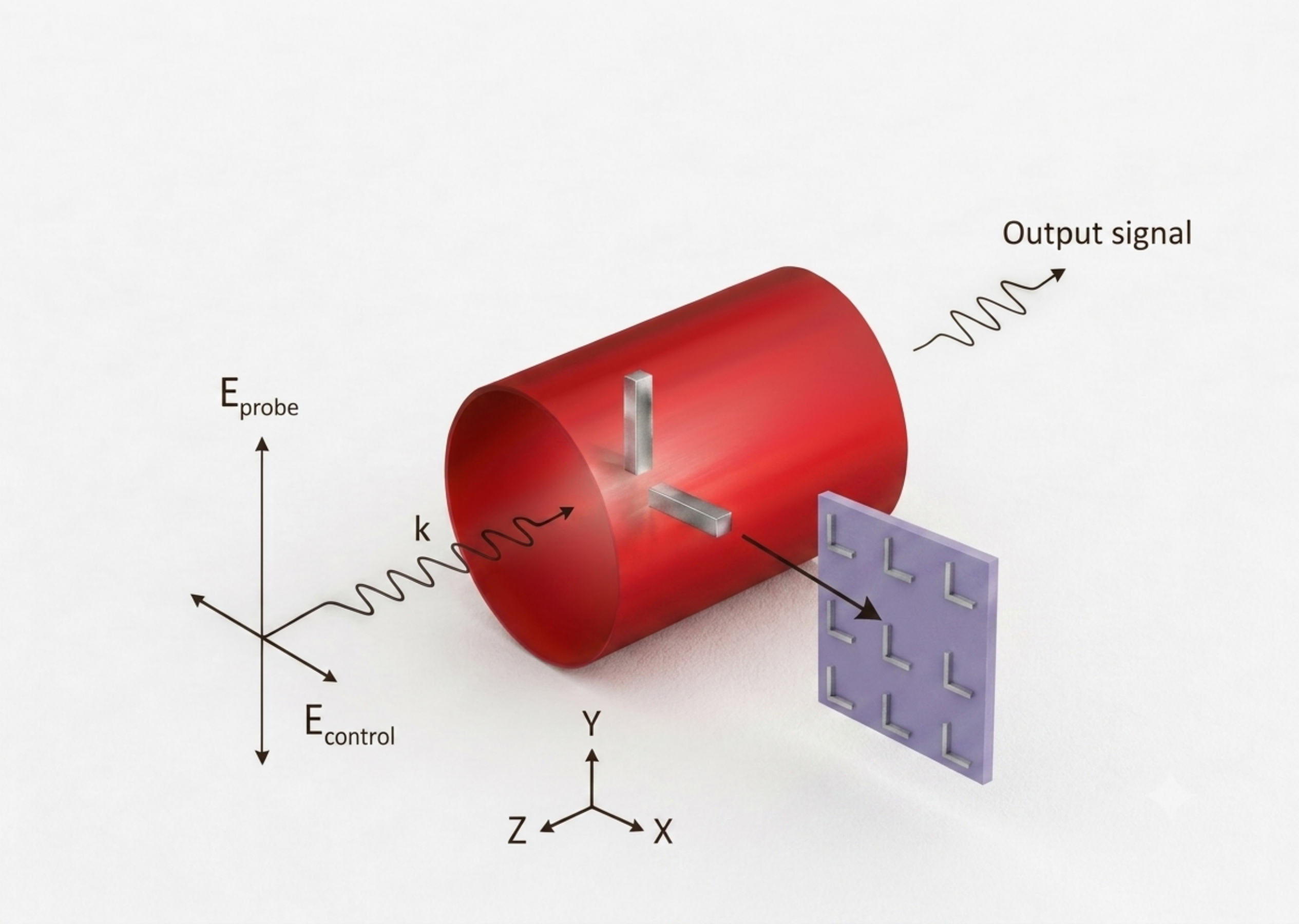}
    \caption{Orthogonal plasmonic nanorod dimer. The probe field $E_1\equiv E_{\rm probe}$ drives one nanorod, and a mutually coherent perpendicular control field $E_2\equiv E_{\rm ctrl}$ drives the other. The control amplitude and phase reshape the probe susceptibility and thereby tune the SHG-induced generation of quantum correlations. The experiment of Ref.~\cite{dhama2022} is performed on a periodically patterned surface covered with such nanorod dimers }
    \label{fig:model}
\end{figure}

The principle of the device is sketched in Fig.~\ref{fig:model}. The probe nanorod supports a fundamental plasmon mode $\hat a_1$ and an effective second-harmonic mode $\hat a_2$; the control nanorod supports $\hat b_1$ and $\hat b_2$~\cite{celebrano2015}. Near-field coupling between the two orthogonal fundamental modes produces a controllable driven susceptibility for the probe polarization. In the classical coupled-oscillator limit~\cite{panahpour2019,stockmann,bookchapter} this \textit{effective} susceptibility can be written as
\begin{equation}
    \chi_{\rm \scriptscriptstyle EIR}(\omega,\Ccal,\phi)
    =\chi_0\frac{\delta_b+\eta\Ccal e^{-\ii\phi}}
    {\delta_a\delta_b-\eta^2},
    \label{eq:eir_susceptibility}
\end{equation}
where $\delta_j=\omega_j^2-\omega^2-\ii\gamma_j\omega$ is the complex detuning of nanorod $j$, $\eta$ is the classical near-field coupling in the oscillator equation, $\Ccal=|E_{\rm ctrl}/E_{\rm probe}|$, and $\phi$ is the control phase. Equation~\eqref{eq:eir_susceptibility} makes the switching principle explicit: changing either $\Ccal$ or $\phi$ alters both the real and imaginary parts of the probe response. Negative $\mathrm{Im}[\chi_{\rm \scriptscriptstyle EIR}]$ in this driven susceptibility is not material gain; it denotes coherent energy redistribution between two phase-correlated polarization channels~${}^{\ref{two_optical_modes}}$~\footnote{The control field $E_{\rm ctrl}$ establishes a coherent background polarization in the $\hat{a}_1$ nanorod that is independent of the probe field,  $P_{\rm ctrl}(\omega)=\eta E_{\rm ctrl} e^{-\ii\phi} / (\delta_a\delta_b-\eta^2)$. The total polarization $P(\omega)=P_{\rm ctrl}(\omega) + \chi_0 E_{\rm probe}(\omega) $ enters the constitutive relation  $D(\omega)=P(\omega)+ \epsilon_0 E_{\rm probe}(\omega)$. Consequently, the probe propagates in the presence of a control-induced coherent polarization background, which modifies its effective refractive index and produces the large phase shifts characteristic of EIR, without invoking material gain. }

{\it Quantum model.}---
The quantum dynamics are modeled by four lossy bosonic modes. In the rotating frame, the Hamiltonian is
\begin{align}
\frac{\hat H}{\hbar}={}&\sum_{s=a,b}\sum_{m=1,2}\Delta_{sm}\hat s_m^\dagger\hat s_m
+g\left(\hat a_1^\dagger\hat b_1+\hat b_1^\dagger\hat a_1\right) \nonumber\\
&+\ii\left(\epsilon_p\hat a_1^\dagger-\epsilon_p^*\hat a_1+\epsilon_c e^{-\ii\phi}\hat b_1^\dagger-\epsilon_c^*e^{\ii\phi}\hat b_1\right) \nonumber\\
&+\sum_{s=a,b}\chi_s^{(2)}\left(\hat s_2^\dagger\hat s_1^2+\hat s_1^{\dagger 2}\hat s_2\right),
\label{eq:hamiltonian}
\end{align}
where $\hat s_m$ denotes $\hat a_m$ or $\hat b_m$, $g$ is the fundamental-mode hopping rate, $\epsilon_p$ and $\epsilon_c$ are proportional to the incident probe and control amplitudes, and $\Delta_{s1}=\omega_{s1}-\omega$, $\Delta_{s2}=\omega_{s2}-2\omega$.  The relative phase $\phi$ is also retained explicitly because it is the experimentally accessible parameter that controls the plasmonic EIR response in addition to the $|E_{\rm ctrl}/E_{\rm probe}|$ ratio.

The key quantum-control mechanism is governed by the SHG interaction term $\hat{a}_2^\dagger \hat{a}_1^2+\mathrm{H.c.}$. Since the probe-mode polarization $\alpha_1=\langle \hat{a}_1\rangle$ scales with the effective susceptibility $\chi_{\rm \scriptscriptstyle EIR}$, it can be controlled through the susceptibility $\chi_{\rm \scriptscriptstyle EIR}(\omega,\mathcal{C},\phi)$ of Eq.~\eqref{eq:eir_susceptibility} by varying the field-amplitude ratio $|E_{\rm ctrl}/E_{\rm probe}|$ and the relative phase $\phi$. Consequently, the control field regulates the strength of the SHG interaction: larger values of $\alpha_1$ lead to more efficient frequency conversion and a larger second-harmonic amplitude $\langle \hat{a}_2\rangle$.


Langevin equations~\cite{gardiner2004,stockmann,tasgin2023} can be written in the form
\begin{align}
	\dot{\hat a}_1 &= -\left(\gamma_{a1}+\ii\Delta_{a1}\right)\hat a_1
	-\ii g\hat b_1
	-\ii 2\chi_a^{(2)}\hat a_1^\dagger\hat a_2
	+\epsilon_p+\sqrt{2\gamma_{a1}}\,\hat a_{1,\mathrm{in}}, \\
	\dot{\hat a}_2 &= -\left(\gamma_{a2}+\ii\Delta_{a2}\right)\hat a_2
	-\ii\chi_a^{(2)}\hat a_1^2
	+\sqrt{2\gamma_{a2}}\,\hat a_{2,\mathrm{in}}, \\
	\dot{\hat b}_1 &= -\left(\gamma_{b1}+\ii\Delta_{b1}\right)\hat b_1
	-\ii g\hat a_1
	-\ii 2\chi_b^{(2)}\hat b_1^\dagger\hat b_2
	+\epsilon_c e^{-\ii\phi}+\sqrt{2\gamma_{b1}}\,\hat b_{1,\mathrm{in}}, \\
	\dot{\hat b}_2 &= -\left(\gamma_{b2}+\ii\Delta_{b2}\right)\hat b_2
	-\ii\chi_b^{(2)}\hat b_1^2
	+\sqrt{2\gamma_{b2}}\,\hat b_{2,\mathrm{in}}.
	\label{eq:langevin}
\end{align}
 The input operators $\hat a_{j,\mathrm{in}}$ and $\hat b_{j,\mathrm{in}}$ describe vacuum noise entering each lossy channel~\cite{gardiner2004,tasgin2023}. The effective nonlinear coefficients $\chi_j^{(2)}$ characterize the surface- and geometry-induced second-harmonic response of the localized plasmonic modes and should therefore be regarded as phenomenological mode coefficients rather than as the bulk second-order susceptibility of a centrosymmetric metal. Additional SHG contributions may arise from nonlinear molecules adsorbed on or coating the nanorod surfaces. If only the $\hat{a}_1$ nanorod is functionalized in this way, the effective nonlinearities satisfy $\chi_a^{(2)}\gg\chi_b^{(2)}$, allowing the second-harmonic coupling to the $\hat{b}_2$ mode to be neglected, as assumed throughout this work.
 $\gamma_{sj}$ are the damping rates of the plasmon modes.

We decompose each operator into its steady coherent amplitude and a zero-mean fluctuation,
\begin{equation}
	\hat a_j=\alpha_j+\delta\hat a_j,\qquad
	\hat b_j=\beta_j+\delta\hat b_j .
\end{equation}
The mean fields are obtained by setting the time derivatives in Eq.~\eqref{eq:langevin} to zero and dropping noise.  Retaining only terms linear in the fluctuations~\cite{genes2008,tasgin2023} gives
\begin{widetext}
	\begin{align}
		\delta\dot{\hat a}_1 &= -\left(\Gamma_{a1}+\ii\Delta_{a1}\right)\delta\hat a_1
		-\ii g\delta\hat b_1
		-\ii 2\chi_a^{(2)}\left(\alpha_2\delta\hat a_1^\dagger+\alpha_1^*\delta\hat a_2\right)
		+\sqrt{2\Gamma_{a1}}\,\hat a_{1,\mathrm{in}}, \\
		\delta\dot{\hat a}_2 &= -\left(\Gamma_{a2}+\ii\Delta_{a2}\right)\delta\hat a_2
		-\ii 2\chi_a^{(2)}\alpha_1\delta\hat a_1
		+\sqrt{2\Gamma_{a2}}\,\hat a_{2,\mathrm{in}}, \\
		\delta\dot{\hat b}_1 &= -\left(\Gamma_{b1}+\ii\Delta_{b1}\right)\delta\hat b_1
		-\ii g\delta\hat a_1
		-\ii 2\chi_b^{(2)}\left(\beta_2\delta\hat b_1^\dagger+\beta_1^*\delta\hat b_2\right)
		+\sqrt{2\Gamma_{b1}}\,\hat b_{1,\mathrm{in}}, \\
		\delta\dot{\hat b}_2 &= -\left(\Gamma_{b2}+\ii\Delta_{b2}\right)\delta\hat b_2
		-\ii 2\chi_b^{(2)}\beta_1\delta\hat b_1
		+\sqrt{2\Gamma_{b2}}\,\hat b_{2,\mathrm{in}}.
		\label{eq:linearized}
	\end{align}
\end{widetext}
The fluctuation decay rates $\Gamma_{sj}$ have been written separately from the classical intensity decay rates $\gamma_{sj}$ to allow for possible additional decoherence channels~\cite{huckPRL2009,di2012quantum}.  In the calculations below the linearized state is retained only in stable regions, i.e., all eigenvalues of the drift matrix have negative real parts.

Retaining terms linear in the fluctuations gives parametric contribution proportional to $\chi_a^{(2)}\bar a_2\delta\hat a_1^\dagger$. Thus the optical control field first determines the steady fundamental fields through Eq.~\eqref{eq:eir_susceptibility}; those fields determine the second-harmonic amplitude $\alpha_2$; and the latter set the squeezing and entangling rates.

The linearized Langevin equations have the standard form
\begin{equation}
    \dot{\bm u}=A\bm u+\bm n(t),
    \label{eq:drift}
\end{equation}
where
\begin{equation}
\bm u=(\delta x_{a1},\delta p_{a1},\delta x_{a2},\delta p_{a2},
\delta x_{b1},\delta p_{b1},\delta x_{b2},\delta p_{b2})^T,
\end{equation}
with $\delta x_o=(\delta\hat o+\delta\hat o^\dagger)/\sqrt2$ and $\delta p_o=(\delta\hat o-\delta\hat o^\dagger)/(\ii\sqrt2)$. Only stable operating points, for which all eigenvalues of $A$ have negative real parts, are retained. For vacuum input noise the steady covariance matrix satisfies
\begin{equation}
    AV+VA^T=-D,
    \label{eq:lyapunov}
\end{equation}
where $D$ follows from the loss channels \cite{gardiner1985,gardiner2004,clerk2010,genes2008,bartels1972,horn2012}. Further details on the construction of $A$ and $D$ are provided in the Supplementary Material~(SM). From the reduced covariance matrix of the two fundamental modes, we quantify entanglement by the logarithmic negativity
\begin{equation}
    \EN=\max[0,-\log_2(2\nu_-)],
    \label{eq:EN}
\end{equation}
where $\nu_-$ is the smaller symplectic eigenvalue of the partially transposed covariance matrix and the vacuum variance is $1/2$ \cite{simon2000,duan2000,adesso2007,vidal2002,plenio2005}. We also report the single-mode nonclassicality $\NSM$, obtained by mixing a fundamental mode with vacuum on a balanced beam splitter and evaluating the logarithmic negativity between the two outputs \cite{asboth2005,kim2002,tasgin2020}. Details on the calculation of $\EN$ and $\NSM$ are given in the Supplementary Material.

\begin{figure}[t]
\centering
\includegraphics[width=0.93\columnwidth]{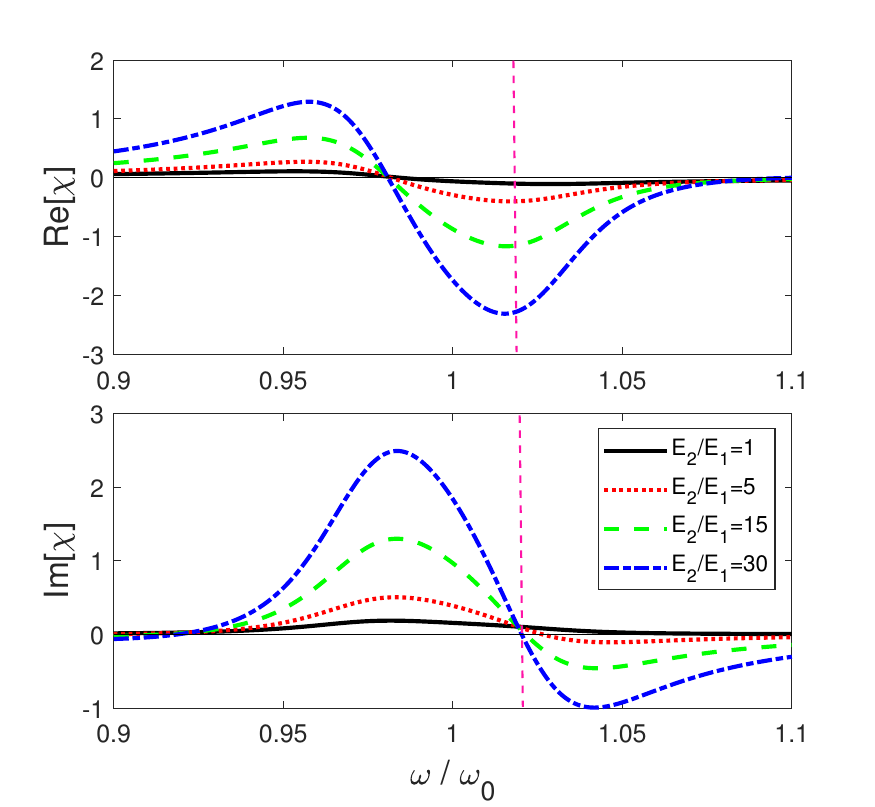}
\caption{Effective susceptibility of the coherently driven plasmonic system. The control ratio $E_2/E_1$~($E_2\equiv E_{\rm ctrl}$ and $E_1\equiv E_{\rm probe}$) strongly enhances the dispersive response and reshapes the absorptive response. Around $\omega\simeq1.02\omega_0$, the effective absorption vanishes while a substantial dispersive response remains.}
\label{fig:linear_response}
\end{figure}

{\it Results.}---
Figure~\ref{fig:linear_response} displays the linear response responsible for the quantum switch. Increasing $E_2/E_1$~($E_2\equiv E_{\rm ctrl}$ and $E_1\equiv E_{\rm probe}$) from 1 to 30 amplifies the dispersive swing in $\mathrm{Re}\,\chi$ and drives $\mathrm{Im}\,\chi$ through a near-zero-extinction point on the high-frequency side of the resonance. We choose $\omega\simeq1.02\omega_0$ for the quantum calculation because it combines strong coherent response with strongly suppressed effective absorption. In principle, an analogous control field applied at $2\omega$ could suppress the effective loss of the second-harmonic plasmon as well, provided that the relevant $2\omega$ modes support directional coupling \cite{Bernasconi2018}. The present calculation therefore demonstrates the fundamental-frequency version of a more general absorption-management strategy.

\begin{figure}[t]
\centering
\includegraphics[width=0.93\columnwidth]{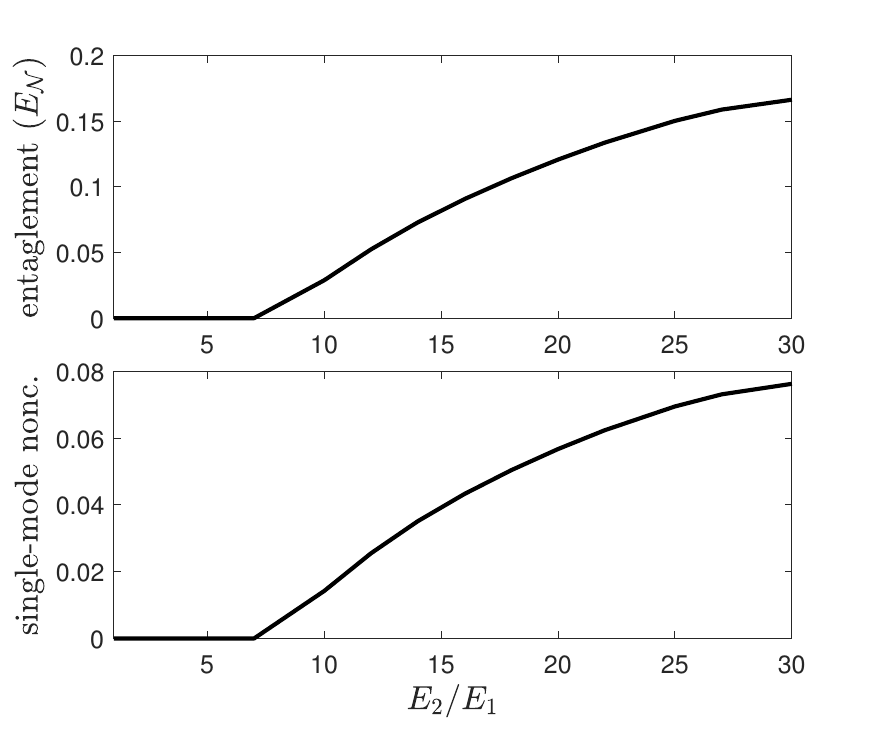}
\caption{Optical control of quantum correlations at $\omega\simeq1.02\omega_0$. The logarithmic negativity $\EN$ between the fundamental plasmon modes and the single-mode nonclassicality $\NSM$ switch on as the control-to-probe ratio $E_2/E_1$ is increased.}
\label{fig:quantum_control}
\end{figure}

The resulting quantum correlations are shown in Fig.~\ref{fig:quantum_control}. At small control ratios the covariance matrix remains separable and classically allowed, because the SHG-induced parametric rate is too weak compared with damping and vacuum noise. Once $E_2/E_1$ exceeds approximately 7 for the displayed parameters, both $\EN$ and $\NSM$ become nonzero. They then increase throughout the plotted range, reaching approximately $\EN\simeq0.17$ and $\NSM\simeq0.08$ at $E_2/E_1=30$. This threshold-like onset is the operational switching action of the nanodimer: the control polarization does not itself supply the entanglement, but it tunes the absorption-suppressed local field that drives the nonlinear parametric interaction.

The simultaneous growth of $\NSM$ and $\EN$ also identifies the resource conversion mechanism. SHG first produces nonclassical quadrature fluctuations locally; the inter-rod coupling then converts part of that single-mode nonclassicality into bipartite continuous-variable entanglement between the fundamental plasmon modes. The plotted values are proof-of-principle rather than optimized limits. Larger correlations should be accessible by optimizing modal overlap, out-coupling, Ohmic and radiative damping, the driving frequency, the control phase, and the control amplitude. The relative phase $\phi$ is particularly important: in Eq.~\eqref{eq:eir_susceptibility} it selects the interference condition that determines whether the response is predominantly dispersively enhanced or absorptively suppressed, while in Eq.~\eqref{eq:hamiltonian} it fixes the complex steady fields and hence the squeezed quadratures.

{\it Conclusion.}---
We have proposed a compact plasmonic nanodimer that functions as an all-optical switch for continuous-variable entanglement. The central advance is the merger of two normally separate ideas: plasmonic refractive-index enhancement, which can move a driven nanorod response to a near-zero-absorption operating point, and nonlinear SHG, which converts that controlled local field into squeezing and entanglement. The same optical control ratio that suppresses effective extinction also turns on logarithmic negativity and single-mode nonclassicality, providing a direct route to reconfigurable nonclassical-light generation in a structure far below the diffraction limit. This addresses a key limitation of metallic quantum-plasmonic devices---loss---without sacrificing the compactness needed for integrated quantum circuits.

Furthermore, unlike conventional all-optical switches based on nonlinear control processes~\cite{nonlinear_switching}, the present scheme operates through control-induced modification of the linear optical susceptibility. The switching action is therefore decoupled from the control nonlinearity itself, avoiding additional nonlinear signatures and the parasitic effects often associated with strong nonlinear switching mechanisms. The SHG process remains solely responsible for generating the nonclassical states, while the control field determines their strength through linear susceptibility engineering.

Furthermore, the switching action is achieved without introducing an additional nonlinear control process and is therefore effectively fingerprint-free, in the sense that the control mechanism does not leave extra nonlinear signatures on the generated quantum state. Instead, the control field reshapes the linear optical response, while the existing SHG mechanism translates this modification into changes in squeezing and entanglement.

The results establish a principle rather than a fully optimized device design. Their significance is that absorption management, tunability, and quantum-resource generation are achieved by the same phase-coherent optical control channel. Future designs can extend the strategy by adding a second control field at $2\omega$, optimizing the nanorod geometry and nonlinear modal overlap, and mapping $\EN(\omega,\Ccal,\phi)$ and $\NSM(\omega,\Ccal,\phi)$ over the full control space. Such developments would turn the present proof of principle into a quantitatively predictive platform for nanoscale, switchable, and absorption-suppressed quantum light sources.

We gratefully acknowledge Eray T\"uz\"un for his assistance with the numerical simulations.

\bibliographystyle{apsrev4-2}
\bibliography{bibliography}

\end{document}